\documentclass[aps,nofootinbib,reprint,nobibnotes,showpacs,superscriptaddress]{revtex4-1} 

\usepackage[english]{babel}
 \addto\captionsenglish{
                       }

\usepackage{graphicx}
 \graphicspath{{graphics/}}

\usepackage[caption=false]{subfig}
\usepackage[export]{adjustbox}

\usepackage{amsmath}
\usepackage{amssymb}
\usepackage[version=3]{mhchem}
\usepackage{mathrsfs}
\usepackage{bm}

\usepackage{upgreek}

\usepackage[hidelinks,linktocpage]{hyperref}
\hypersetup{
            colorlinks,
            linkcolor={red!70!black},
            citecolor={green!70!black},
            urlcolor={blue!70!black}
           }

\usepackage[usenames,dvipsnames]{xcolor}
\usepackage{comment} 

\usepackage{xspace}

\newcommand{\al}{\ensuremath{\alpha}\xspace}
\newcommand{\bt}{\ensuremath{\beta}\xspace}
\newcommand{\gm}{\ensuremath{\gamma}\xspace}

\newcommand{\RaP}{\ce{^{228}Ra}\xspace}
\newcommand{\RaS}{\ce{^{224}Ra}\xspace}

\newcommand{\BiPo}{\ce{Bi-Po}\xspace}
\newcommand{\Ea}{\ensuremath{E_\al}\xspace}
\newcommand{\Qb}{\ensuremath{Q_\bt}\xspace}

\newcommand{\muA}{\ensuremath{\upmu\text{A}}\xspace}
\newcommand{\mum}{\ensuremath{\upmu\text{m}}\xspace}
\newcommand{\mus}{\ensuremath{\upmu\text{s}}\xspace}
\newcommand{\ImuC}{\ensuremath{\upmu\text{C}^{-1}}\xspace}
\newcommand{\Is}{\ensuremath{\text{s}^{-1}}\xspace}

\newcommand{\sI}{\ensuremath{S1}\xspace}
\newcommand{\sII}{\ensuremath{S2}\xspace}
\newcommand{\sA}{\ensuremath{S3}\xspace}
\newcommand{\sB}{\ensuremath{S4}\xspace}

\newcommand{\hS}[1]{{\hspace{#1 pt}}}

\begin{document}

 \title{Production of monochromatic \RaP \boldmath{\al}-sources for detector characterization}

 \author{M.\ Biassoni}
  \affiliation{INFN Sezione di Milano-Bicocca, I-20126 Milano, Italy}
 \author{C.\ Brofferio}
  \email{chiara.brofferio@unimib.it}
  \affiliation{University of Milano-Bicocca, I-20126 Milano, Italy}
  \affiliation{INFN Sezione di Milano-Bicocca, I-20126 Milano, Italy}
 \author{S.\ Dell'Oro}
  \affiliation{University of Milano-Bicocca, I-20126 Milano, Italy}
  \affiliation{INFN Sezione di Milano-Bicocca, I-20126 Milano, Italy}
 \author{L. Gironi}
  \affiliation{University of Milano-Bicocca, I-20126 Milano, Italy}
  \affiliation{INFN Sezione di Milano-Bicocca, I-20126 Milano, Italy}
 \author{M.\ Nastasi}
  \affiliation{University of Milano-Bicocca, I-20126 Milano, Italy}
  \affiliation{INFN Sezione di Milano-Bicocca, I-20126 Milano, Italy}
 \author{M.\ Sisti}
  \affiliation{INFN Sezione di Milano-Bicocca, I-20126 Milano, Italy}

 \date{\today}
   
 \begin{abstract}
  The response of particle detectors to different types of radiation is not necessarily identical and, in some cases, neglecting this behavior can lead to a misinterpretation of the acquired data.
  While commercial radioactive sources are in general suitable to investigate the response to \bt's and \gm's, in the case of \al's the need for custom-made sources arises from the intrinsic
  properties of \al radiation, which imposes that the emitter directly faces the detector.
  In this work, we show how to flexibly produce \al sources to be employed in multiple studies of detector characterization.
  These are obtained starting from a set of primary sources obtained from the collection of radioactive \RaP ions at the ISOLDE facility at CERN.
  We illustrate the potential of this technique with practical cases of application to scintillators and bolometric detectors and examples of the results obtained so far.
  \\[+9pt]
  Published on: Eur.\ Phys.\ J.\ Plus {\bf 137}, 1309 (2022) \hfill DOI: \href{https://doi.org/10.1140/epjp/s13360-022-03519-4}{10.1140/epjp/s13360-022-03519-4}
 \end{abstract}

 \maketitle

 \section{Introduction}

 The signal shape and evolution, the dependence of the signal amplitude on the energy deposition and the energy resolution in a particle detector
 can exhibit measurable differences for \al, \bt or \gm radiation.
 A quantitative assessment of these effects might be needed during multiple phases of an experiment cycle, from calibration, in order to correct for shifts in the peak positions,
 to data analysis, where such knowledge could be beneficial for the development of effective rejection techniques to improve the signal-to-noise ratio.
 
 In general, to study the response to \bt- or \gm-particle interactions of materials employed as scintillators or bolometers (a.\,k.\,a.\ cryogenic macro-calorimeters),
 a detector can simply be exposed to an external source placed in its proximity.
 In the case of \al's, however, the source needs to directly face the detector without interposed material, therefore geometrical constraints often limit (or even preclude) the use of commercial elements.
 Furthermore, since multiple active spots can be needed during the same measurement (up to one per individual `sub-detector', e.\,g.\ a crystal inside an array),
 the source availability can become a relevant issue as well; in these cases, the preparation of custom \al sources emerges almost as a mandatory solution.
 
 Producing a suitable \al source is far from being trivial. Apart from the need of actually obtaining an \al-emitting nuclide or a progenitor, standard methods such as diffusion and deposition
 (either chemical or electro) should be avoided because they are unable to create a sufficiently thin active layer:
 the resulting self-absorption effects would translate into a line-broadening that in turn would prevent a full exploitation of the good energy resolution of solid-state detectors.
 At the same time, for some applications a source should emit three or more energy quanta in order to allow for a more robust calibration. 
 These constraints significantly reduce the number of available possibilities.

 In this work, we propose the collection of radioactive \RaP from an ion beam as an effective technique for the preparation of home-made \al sources, being \RaP the progenitor
 of a cascade of \al-emitter nuclides.
 This set of \emph{primary} sources can in fact be utilized to generate \emph{secondary} sources by collecting the still-radioactive recoiling nuclei of the \RaP daughters
 (those which \al-decay) on a different target.
 Once implanted, these nuclides become the actual \al emitters of the secondary source.
 We describe the primary-source production and the characterization of a few samples of primary and secondary sources tested in multiple runs with \al spectrometers and bolometric detectors.
 Finally, we illustrate examples of potential applications in detector characterization.

 \section{Source description}

 \begin{figure*}
  \captionsetup[subtable]{labelformat=empty}
  \captionsetup[subfigure]{labelformat=empty}
  \centering
  \subfloat[]
           {\includegraphics[width=0.9\columnwidth,valign=c]{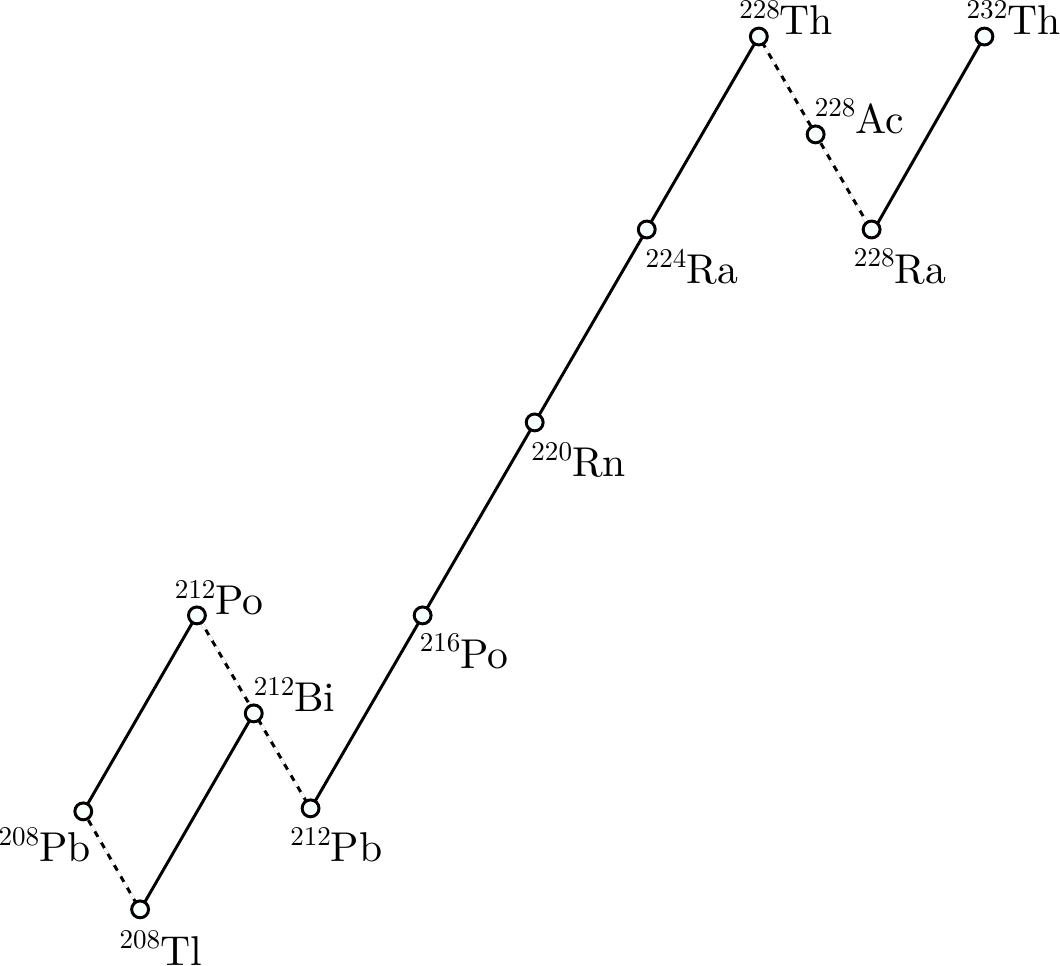}}
  \qquad
  \subfloat[]
           {\adjustbox{valign=c}
                      {
                       \begin{tabular}{l r r r r r}
                        Nuclide          &Half-life        &\multicolumn{2}{r}{Decay (\%)}              &Q-value              &\Ea                 \\
                        \hspace{20pt}    &\hspace{40pt}    &\hspace{25pt}\al       &\hspace{25pt}\bt    &\hspace{20pt} [keV]  &\hspace{20pt} [keV] \\
                        \cline{1-6}                                                                                                                \\[-5pt]
                        \ce{^{232}Th}    &$14.02$ Gyr      &$78.9$    &                 &$4062$                                   &$4011$          \\
                                         &                 &$21.0$    &                 &\textquotesingle\textquotesingle         &$3948$          \\
                        \multicolumn{2}{l}{\underline{\textit{Primary source}}}                                                                    \\[5pt]
                        \ce{^{228}Ra}    &$5.75$ yr        &          &$100$            &$45.8$                                   &                \\
                        \ce{^{228}Ac}    &$6.15$ h         &          &$100$            &$2124$                                   &                \\
                        \ce{^{228}Th}    &$1.91$ yr        &$73.2$    &                 &$5520$                                   &$5423$          \\
                                         &                 &$26.2$    &                 &\textquotesingle\textquotesingle         &$5340$          \\
                        \multicolumn{2}{l}{\underline{\textit{Secondary source}}}                                                                  \\[5pt]
                        \ce{^{224}Ra}    &$3.63$ d         &$94.7$    &                 &$5789$                                   &$5686$          \\
                                         &                 &$5.3$     &                 &\textquotesingle\textquotesingle         &$5449$          \\
                        \ce{^{220}Rn}    &$55.8$ s         &$99.9$    &                 &$6405$                                   &$6288$          \\
                        \ce{^{216}Po}    &$0.15$ s         &$100$     &                 &$6907$                                   &$6779$          \\
                        \ce{^{212}Pb}    &$10.64$ h        &          &$100$            &$574$                                    &                \\
                        \ce{^{212}Bi}    &$60.54$ min      &$25.1$    &                 &$6207$                                   &$6051$          \\
                                         &                 &$9.7$     &                 &\textquotesingle\textquotesingle         &$6090$          \\
                                         &                 &          &$64.1$           &$2254$                                   &                \\
                        \ce{^{212}Po}    &$300$ ns         &$100$     &                 &$8954$                                   &$8785$          \\
                        \ce{^{208}Tl}    &$3.06$ min       &          &$100$            &$5001$                                   &                \\
                        \ce{^{208}Pb}    &stable           &          &                 &                                         &
                       \end{tabular}
                      }
           }
  \caption{The $4n$ chain of \ce{^{232}Th}.
           \textit{(Left)} Decay scheme in a $(N,Z)$ plane. The solid lines represent \al decays ($N \to N-2$, $Z \to Z-2$), the dashed lines represent the $\bt^-$ decays  ($N \to N-1$, $Z \to Z+1$).
           \textit{(Right)} Main decay channels for the various nuclides of the series; since only values greater than $1\%$ are listed, the overall probability per transition can be lower than $100\%$.
           The uncertainties on the numerical values are smaller or comparable to the quoted precision~\cite{BIPM_tabRadNuc} and have not been reported.
           The precursors of the primary and secondary sources are highlighted.
          }
  \label{fig:chain}
 \end{figure*}

 \RaP is the daughter of \ce{^{232}Th}, the progenitor of the $4n$ natural-radioactivity chain (Fig.~\ref{fig:chain}).
 After implantation, \RaP begins to \bt-decay to \ce{^{228}Th} (passing through \ce{^{228}Ac}) so that the two populations will eventually reach equilibrium after some years, given the relative
 half-lives.
 Down the chain, instead, the source `evolution' over time becomes more complex to describe since the equilibrium is potentially broken at the level of any following \al decay.
 In fact, given the very shallow implantation of the radioactive nuclides, each \al decay from the source can actually result in the emission of the recoiling nucleus rather than the \ce{He^{2+}} ion,
 thus preventing the continuation of the chain from that specific decay on.
 This behavior is at the base of the production of the secondary sources, which are obtained by collecting these ejected recoiling nuclei, in turn \al-emitters themselves.

 \RaP has a half-life of almost 6 years, which ensures a very convenient lifetime for a source, both for a direct use in calibration and for the production of secondary sources.
 The maximum overall activity of the primary source is reached about $5$ years after the production, since the decrease of \RaP is gradually compensated by the increase of \ce{^{228}Th}
 and all its daughters.

 Up to 7 \al peaks are expected to show up in the \RaS-source spectrum, while two additional peaks from \ce{^{228}Th} are expected for the primary source.

 \subsection{Primary-source production}

 \begin{table*}[t]
  \caption{
           Source details and specifications of the ion-collection process at ISOLDE.
           The activity of \RaP after implantation has been reconstructed from the measurement of the \gm peaks of \ce{^{228}Ac}; the error accounts for both the statistical and systematical components.
           The different activities of \sI and \sII are due to an asymmetric positioning of the two \ce{Ta} foils with respect to the beam during the ion-collection process.
          }
  \vspace{5pt}
  \centering
  \begin{tabular}{l r r r r r r r r}
   Source  &\hS{5} Production &\hS{5} Surface $[\text{mm}^2]$ &\hS{5} Target &\hS{5} Src Ionizer &\hS{5} Avg $I_\mathrm{p}$ [\muA] &\hS{5} $\Delta$V [kV] &\hS{5} $t_\mathrm{coll.}$ [h] &\hS{20} $A_0$ (\RaP) [Bq]   \\
   \cline{1-9}                                                                                                                                                                                         \\
   \sI     &Jun 2016          &$5.0\times 8.3$                &\ce{UC_x}     &\ce{W}         &$0.5$             &$20$             &$4.8$        &$248 \pm  \hphantom{0}8$           \\
   \sII    &Jun 2016          &$4.2\times 8.3$                &\ce{UC_x}     &\ce{W}         &$0.5$             &$20$             &$4.8$        &$\hphantom{0}74 \pm \hphantom{0}2$ \\
   \sA     &Sep 2018          &$7.8\times 9.8$                &\ce{UC_x}     &\ce{Ta}        &$0.7$             &$20$             &$18.6$       &$1031 \pm 32$                      \\
   \sB     &Sep 2018          &$7.8\times 9.8$                &\ce{UC_x}     &\ce{Ta}        &$0.7$             &$20$             &$18.6$       &$1153 \pm 35$                      \\
  \end{tabular}
  \label{tab:sources}
 \end{table*}

 The \RaP sources have been produced at the Isotope mass Separator On-Line DEvice (ISOLDE) facility at CERN~\cite{Catherall:2017kbr} during two irradiation campaigns in 2016 and 2018.
 At ISOLDE, a fixed target is bombarded with a pulsed proton beam of $1.4$ GeV and intensity up to 2 \muA, which produces a large number of spallation, fragmentation and ﬁssion reactions.
 The reaction products are then extracted, ionized, accelerated and separated to provide a variety of radioactive-ion beams.

 ISOLDE is able to guarantee a \RaP yield up to about $1.8\cdot10^7$ ions \ImuC by using an uranium-carbide (\ce{UC_x}) thick target,
 so that a source with an activity of a few kBq could be obtained with a day-long irradiation.
 For our collection, we could actually rely on a proton current between $0.5$ and $2$ \muA for most of the time,
 while accelerating the \ce{^{228}Ra^+} ions with a $\Delta\text{V}$ of only $20$ kV in order to make the implantation as shallow as possible.
 The implantation rate was thus of about $2\cdot 10^7$ ions \Is.

 We used $200$-\mum \ce{Ta} foils as ion collectors; tantalum is a non-toxic metal highly resistant to corrosion, which makes it ideal as target.
 We produced four \RaP sources, two during the first campaign and two during the second one (Table~\ref{tab:sources}). 

 \subsection{Secondary sources}
 
 The penetration of the \RaP ions obtained at ISOLDE into our \ce{Ta} foils expected from simulations with SRIM-2013~\cite{Ziegler:2010zz} is of a few nanometers and the same holds
 for the various nuclides along the chain because the \al decays will not significantly push the nuclei further into the source bulk (the \bt's will not shift the nuclei).
 This extremely shallow implantation allows for the escape of the recoiling nucleus whenever an \al is emitted inwards with respect to the foil surface;
 indeed, the nucleus kinetic energy will be comprised between $100$ and $170$ keV for \al's between $5.4$ and $8.8$ MeV,
 i.\,e.\ larger than the kinetic energy provided during the ion-beam collection (about $20$ keV).
 The ejected nuclei can therefore be implanted on a separate target.
 
 In order to produce a secondary source, we face either a small aluminum or copper foil, or a scintillating crystal to the primary source.
 We keep the gap in between smaller than $1$ cm and hold this configuration inside a constantly pumped chamber: the vacuum, which we keep at the level of $(10^{-2}-10^{-3})$ mbar,
 reduces the possibility of slowing down the nucleus by collision with an air molecule.
 
 The actual composition and activity of the newly created source are not trivial to compute because, on the one side, all the nuclides from \RaS down the chain can be implanted; on the other,
 they can as well escape the secondary source during its preparation.
 Anyway, after a few days, a saturation condition is reached since the longest half-life among those of the implanted nuclides (that of \RaS) is shorter than $4$ days (Table in Fig.~\ref{fig:chain}).

 \section{Source characterization}

 The primary sources have been measured with a high-purity germanium (HPGe) detector in order to assess the effectiveness of the \RaP collection.
 The activity of \RaP could be reconstructed by measuring the \gm lines of \ce{^{228}Ac}
 and simulating the system geometry with a Monte Carlo to extract the detection efficiency.
 \ce{^{228}Ac} cannot escape the \ce{Ta} foil, and hence, it is a good proxy for the \RaP content.
 The values obtained at the specific time of the measurement could then be rescaled to infer the original activity of each source at the time of implantation.
 The results are reported in Table~\ref{tab:sources}: as it can be seen, they are of some hundreds Bq, consistent with our expectations given the beam intensity and the irradiation time.

\begin{figure}[b]
  \centering
  \includegraphics[width=\columnwidth]{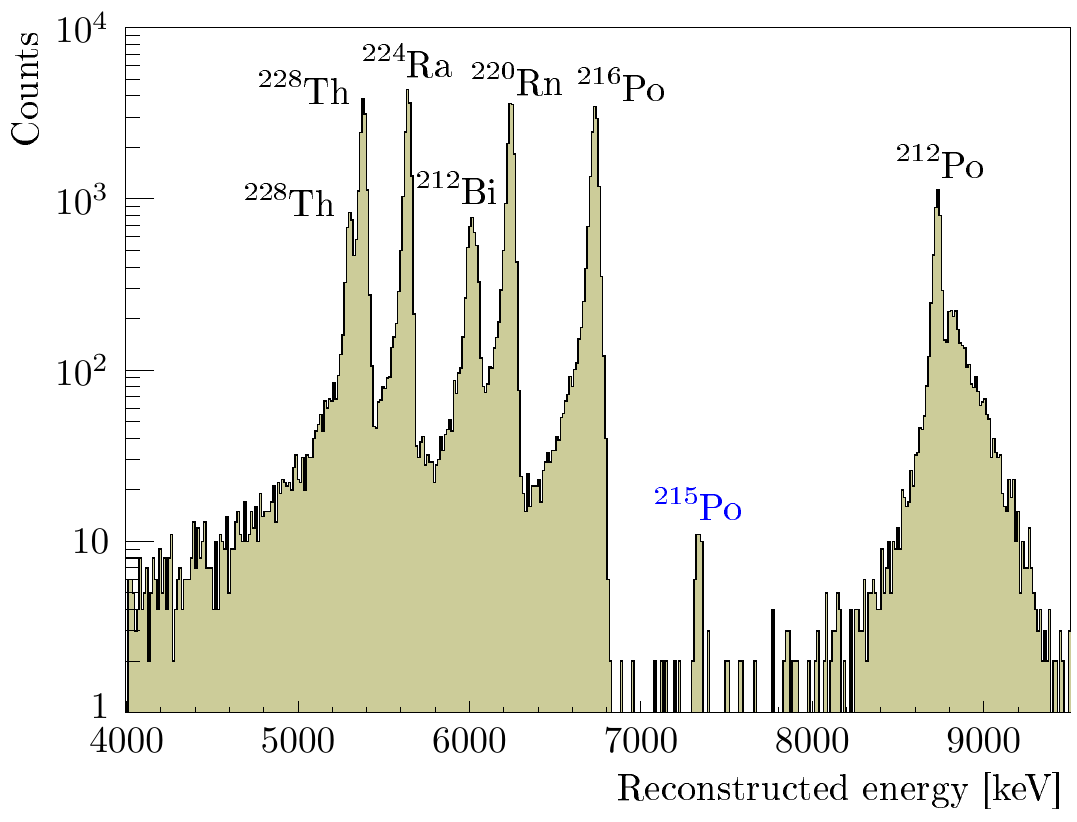}
  \caption{
           Spectrum acquired by exposing a \ce{Si} surface barrier detector to the source \sA\ two years after the implantation at ISOLDE.
           All the 7 peaks from the \RaP decay cascade are clearly visible.
           The sharp \ce{^{215}Po} peak at $7.4$ MeV is due to a contamination of \ce{^{227}Ac} in the source;%
           \protect\footnotemark\ the other peaks from that radioactive chains are covered by the \RaP (and progeny) ones.
          }
 \label{fig:S1_spectrum}
 \end{figure}

 As an independent benchmark, the activity of \sI was also reconstructed by measuring the source \al spectrum with a silicon (\ce{Si}) surface barrier detector.
 The average value from multiple peaks of \RaP was $249 \pm 26$ Bq (the error includes both the statistical and systematical components),
 perfectly compatible with $248 \pm 8$ Bq obtained with the HPGe.%
 \footnote{This cross check was only performed on one source since the measurement with the \al spectrometer contaminates the detector due to the implantation of the recoiling
 nuclei on the \ce{Si} wafer and on the surrounding chamber, so that for about a month ($\sim 10$ half-lives of \RaS) this operates with a higher intrinsic background.}
 Figure~\ref{fig:S1_spectrum} shows the obtained spectrum: 7 peaks from the \RaP decay chain (those with higher decay rate) are clearly identifiable.
 The very shallow implantation results in very sharp peaks, since the emitted \al's leave the source while still carrying the whole energy of the transition.
 The FWHM resolutions are of about $30$ keV; this value is actually limited by the detector intrinsic resolution rather than by the source, still it is already comparable
 with that of the best-performing commercial units.

 \footnotetext{The presence of \ce{^{227}Ac} has been assessed by performing \gm spectroscopy on the source~\cite{Baccolo:2021odk}; its presence is likely due to a parasitic collection
               of \ce{^{227}Ac^+} during the source production at ISOLDE.}
 
 \section{Examples of applications}
 
 The sources produced with the above-described procedure can find multiple useful applications to characterize the response of detectors to \al radiation.
 Here we report a few examples.

 \subsection{Quenching factor}
 
 \begin{figure}[t]
  \centering
  \includegraphics[width=\columnwidth]{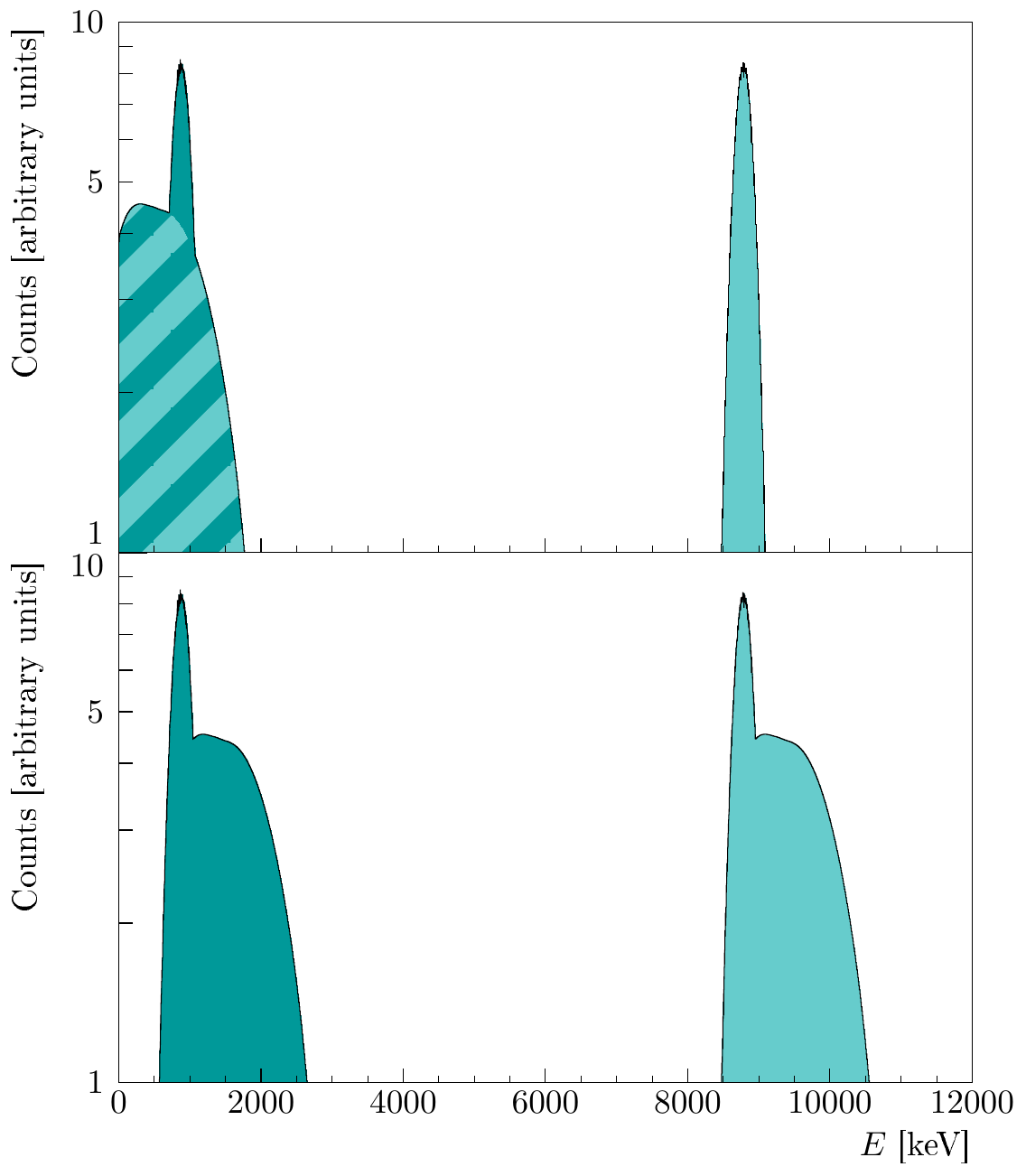}
  \caption{
           Simulated spectra of the \ce{^{232}Th}-chain \BiPo.
           The light spectra would be obtained if \al's and \bt's particles gave the same light yield; the dark spectra illustrate the case of a quenching factor for \al's equal to $0.1$;
           the striped \bt spectrum is unaffected by the quenching.
           The ideal \al peak has a Gaussian shape, while the \bt spectrum is a continuum ending at the Q-value of \ce{^{212}Bi}.
           \emph{(Top)} The detector is sufficiently fast to resolve the decay cascade: \ce{^{212}Bi ->[\bt] ^{212}Po ->[\al] ^{208}Pb}.
           \emph{(Bottom)} The \al and \bt components are convolved. Here, we assume we are detecting both \al and \bt; either could actually escape the detector, hence complicating the
           overall shape of the spectrum.
          }
 \label{fig:BiPo}
 \end{figure}
 
 The different light yield of a scintillator depending on the type of interaction can be represented by a \emph{quenching factor} that multiplies the ratio of the light yield for
 the same deposited energy in \al or \bt/\gm events.
 To quantify this effect, we can take advantage of the final part of the source decay chain, the so-called \BiPo sequence.

 In the \BiPo, the \bt from the decay of the \ce{^{212}Bi} is followed by the \al decay of \ce{^{212}Po} with a half-life of $300$ ns.
 By comparing the reconstructed position of the \al peak ($\Ea=8785$ keV) and that of the Q-value of the \bt decay ($\Qb=2254$ keV) with the corresponding nominal values,
 it is thus possible to extract the conversion between light yield and energy for the two types of interactions.
 Typical values for the quenching factors are in the range ($0.1-0.2$)~\cite{Tretyak:2009sr}, so that the \al peak actually ends up in the \bt/\gm region.
 
 The clear signature of the \BiPo makes the event selection quite straightforward, as rendered in Fig.~\ref{fig:BiPo}.
 The structure showing up in the energy spectrum actually depends on the ability of the detector to resolve the decay cascade.
 For detectors with time resolutions of tens to a few hundreds ns, the \al and \bt components are well separated; for slower detectors the two components are convolved.
 In the former case (upper panel in the figure), the selection of the \al events can be performed by simply cutting on the energy around the corresponding Gaussian peak;
 this in turn allows the tagging of the lower-energy \bt's by constraining the time distance between the two events to be smaller than about $(1-2)$ \mus.
 In the latter case (lower panel in the figure), the event selection can be performed again by cutting on the event energy.
 The \bt spectrum is now shifted by (quenched) \Ea, hence the end-point will be at $\Qb+\Ea$, but we still expect a peak at \Ea due to the events for which only the \al is detected
 while the \bt escapes (simply for geometric reasons).

 \subsection{Self-absorption}
 
 \begin{figure}[t]
  \centering
  \includegraphics[width=\columnwidth]{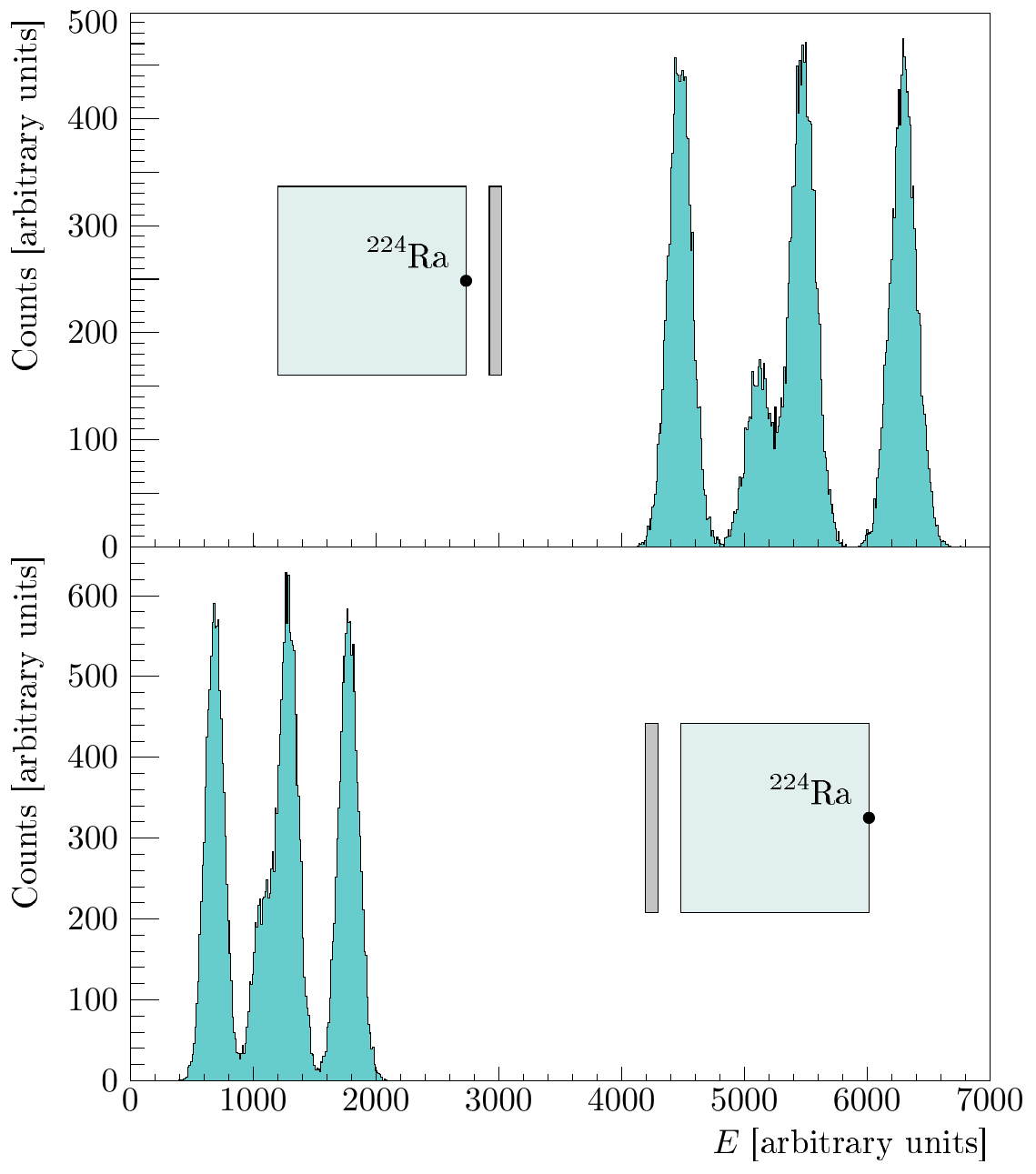}
  \caption{
           Simulated measurement of the light self-absorption with a scintillator with the implanted \al source directly facing a light sensor.
           \emph{(Top)} \RaS is implanted on the side facing the detector: the collected light following an \al decay practically does not cross the crystal.
           \emph{(Bottom)} \RaS is implanted on the opposite side of the detector: the light following an \al decay is collected after crossing the crystal.
           If we assume the same efficiency for the detection of light reaching the light sensor in the two configurations,
           the position of the \al peaks will shift towards lower energies due to the self-absorption of light in the crystal.
           The worsening of the resolution is due to the more unfavorable statistics for the light collection in the latter configuration.
          }
 \label{fig:self_absorption}
 \end{figure}
 
 A scintillator produces photons following a particle interaction; a fraction of the generated light, however, can be re-absorbed while traveling inside the material.
 In order to study the self-absorption of light, one should know where this has been generated and compute the distance travelled inside the material before detection.
 \al particles are ideal in this regard since, being the path between emission and absorption practically negligible, we can identify the light-emission position with that of the \al-emitting nuclide.
 
 The source-production method described in this work allows implanting \RaS directly on scintillators without contaminating them, since \RaS and its daughters are all short-living nuclides:
 after less than a month, the residual activity is in fact totally negligible.
 This technique could thus be applied even to rare and precious specimens without spoiling their radio-purity.

 By exposing the scintillator to a light sensor, so that this first faces the side with the implanted \al source, and then the opposite one,
 the light self-absorption will translate into a shift in the \al-peak positions toward lower energies while passing from the former to the latter configuration (Fig.~\ref{fig:self_absorption}).

 \subsection{Coincidence studies}

 \begin{figure}[t]
  \centering
  \includegraphics[width=\columnwidth]{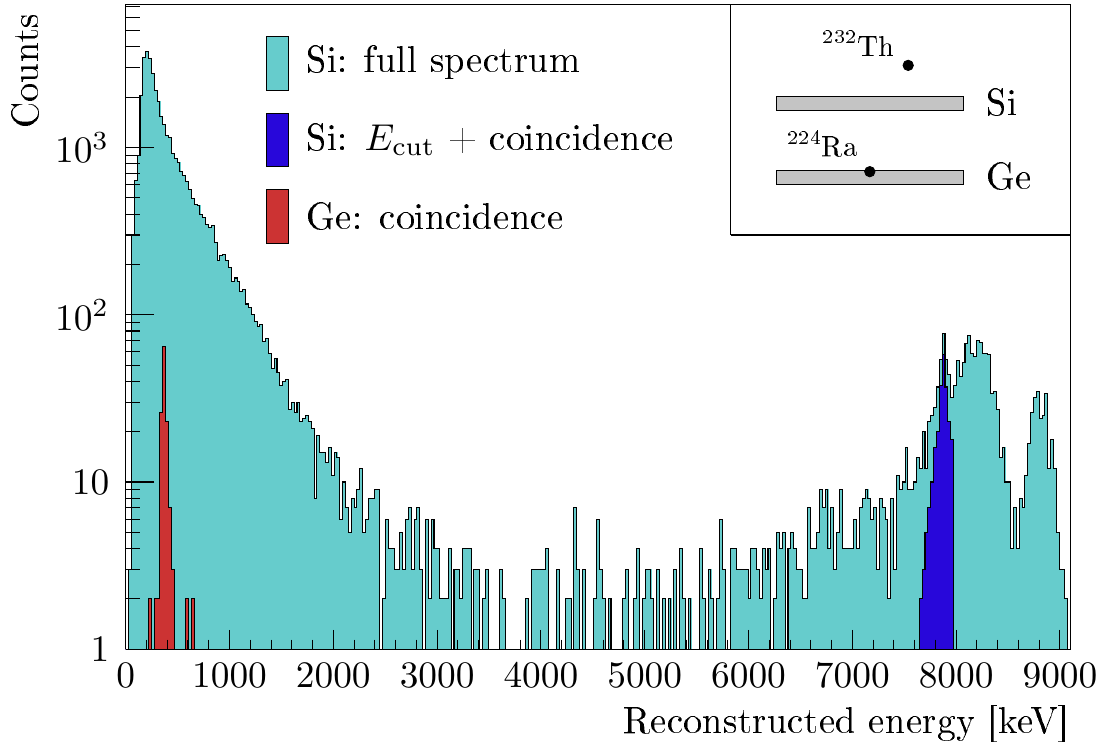}
  \caption{
           Coincidence studies between two bolometric detectors: a \ce{Si} disk faces a \ce{Ge} disk with the implanted \al source, while being also exposed to a liquid \ce{^{232}Th} source.
           The \al events on the \ce{Si} detector from the implanted source are selected by cutting on the energy around the $5.68$-MeV peak of \RaS
           and by imposing a coincidence with the \ce{Ge} detector (blue).
           Despite the higher continuous background due to the liquid source, the corresponding recoiling nuclei show up as a monochromatic peak at low energy on the \ce{Ge} detector (red).
          }
 \label{fig:coinc_spectrum}
 \end{figure}
 
 Understanding the response of scintillators/bolometers to nuclear recoils is a fundamental requirement for experiments based on these technologies,
 which are looking for direct interaction of Dark Matter~\cite{Billard:2021uyg}.  
 This kind of characterization can be performed by using the above-described \RaS sources by exploiting the coincident emission `\al~+~recoiling nucleus',
 where the former actually act as trigger for the identification of the latter.
 In fact, if we expose a detector with implanted \RaS to a second detector, the escaping \al and nuclear recoils will be observed separately on the two
 (with some efficiency related to the geometry of the system);
 by selecting a specific \al peak on either detector, we expect the recoil events to distribute along a monochromatic peak on the other (Fig.~\ref{fig:coinc_spectrum}).
 
 \subsection{Response of bolometric detectors}

 Bolometers can achieve resolutions down to a few per mil at the MeV scale.
 Therefore, in order to fully exploit the high performance of these detectors in studies related to the detector response to \al particles, the energy depositions should be as monochromatic as possible.
 The use of custom \al sources obtained by diffusion or chemical/electro deposition of the radioactive nuclides generally translate into non-Gaussian peaks, with significant tails towards lower
 energies due to the deeper penetration of the \al-emitters inside the source substrate.
 On the contrary, the sources (either primary or secondary) presented in this work are ideal in this regard, by giving fairly sharp peaks.
 
\begin{figure}[t]
  \centering
  \includegraphics[width=\columnwidth]{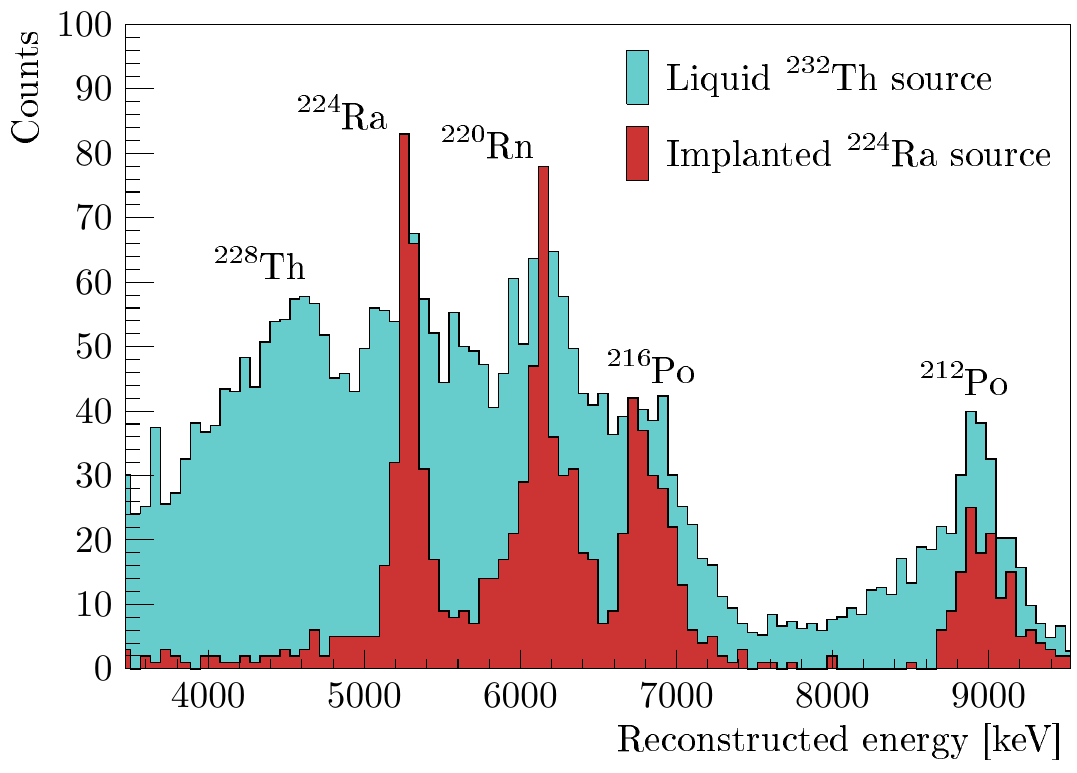}
  \caption{
           Comparison between energy spectra in the \al region for two identical bolometric detectors (\ce{Ge} disks), one exposed to an implanted \RaS source (red),
           the other exposed to a liquid \ce{^{232}Th} source (cyan).
           The peak at $4$ MeV is due to the two unresolved peaks from the decay of \ce{^{232}Th}, hence do not show up in the spectrum of the \RaS source.
          }
 \label{fig:spectra_comparison}
 \end{figure}

 The situation is clearly illustrated in Fig.~\ref{fig:spectra_comparison}, where two identical detectors face a liquid \ce{^{232}Th} source and an implanted \RaS source.
 While in the former case it is hard even to disentangle the various peaks, in the latter case the peaks are monochromatic with FWHM resolutions of about $30$ keV.

 \section{Summary}
 
 We presented an effective technique for the production of home-made \al sources for the characterization of scintillators or bolometric detectors.
 Starting from a set of primary \RaP sources produced by radioactive-ion collection at the ISOLDE facility, we are able in turn to generate multiple secondary sources by implanting on different targets
 all the \al-emitter daughters from \RaS down the \ce{^{232}Th} chain.
 These secondary sources present a tens-of-nm-thick active layer, so that the resulting \al peaks are very sharp and maintain a Gaussian shape, allowing for a full exploitation of the
 good energy resolution of solid state detectors. At the same time, the short half-life of \RaS and its progeny makes it possible to directly implant the radioactive nuclides on
 any detector without spoiling its radiopurity.
 We provided practical examples of studies that can be carried out with the secondary sources, which show the flexibility and potential of this technique.

\section*{Acknowledgments}
 This work makes use of the `Arby' software for Geant4-based Monte Carlo simulations, that has been developed in the framework of the Milano-Bicocca R\&D activities
 and that is maintained by O.\ Cremonesi and S.\ Pozzi.
 The IS604 members M.\,B.\ and M.\,S.\ acknowledge support from the European Union's Horizon 2020 Framework research and innovation programme under Grant Agreement no.\ 654002 (ENSAR2).

\section*{Data availability statement}

 The datasets generated and analyzed during the current study are available from the corresponding author on reasonable request.

 \bibliography{ref}

\end{document}